\documentclass[a4paper, 10pt, conference]{IEEEtran}
\IEEEoverridecommandlockouts

\usepackage{cite}
\usepackage{amsmath,amssymb,amsfonts}
\usepackage{algorithmic}
\usepackage{graphicx}
\usepackage{textcomp}
\usepackage{booktabs}
\usepackage{enumerate}

\makeatletter
\let\MYcaption\@makecaption
\makeatother

\usepackage[font=footnotesize]{subcaption}

\makeatletter
\let\@makecaption\MYcaption
\makeatother

\usepackage{subcaption}

\usepackage{xcolor}

\def\BibTeX{{\rm B\kern-.05em{\sc i\kern-.025em b}\kern-.08em
    T\kern-.1667em\lower.7ex\hbox{E}\kern-.125emX}}
\begin{document}

\title{
    Loc-VAE: Learning Structurally Localized Representation from 3D Brain MR Images\\ for Content-Based Image Retrieval
}

\author{
\IEEEauthorblockN{
    Kei Nishimaki\textsuperscript{\rm 1},
    Kumpei Ikuta\textsuperscript{\rm 1},
    Yuto Onga\textsuperscript{\rm 1},
    Hitoshi Iyatomi\textsuperscript{\rm 1},
    Kenichi Oishi\textsuperscript{\rm 2}
    \\
    for the Alzheimer's Disease Neuroimaging Initiative*
}
\IEEEauthorblockA{
    \textit{\textsuperscript{\rm 1} Department of Applied Informatics, Graduate School of Science and Engineering, Hosei University, Tokyo, Japan}\\
    \textit{\textsuperscript{\rm 2} Department of Radiology and Radiological Science, Johns Hopkins University School of Medicine, Baltimore, USA}\\
    \{kei.nishimaki.1106, kunpei.ikuta, yuuto.onnga.23\}@gmail.com, iyatomi@hosei.ac.jp, koishi2@jhmi.edu
    }
    
\thanks{
    *Data used in preparation of this article were obtained from the Alzheimer's Disease Neuroimaging Initiative (ADNI) database (adni.loni.usc.edu). As such, the investigators within the ADNI contributed to the design and implementation of ADNI and/or provided data but did not participate in analysis or writing of this report. A complete listing of ADNI investigators can be found at: http://adni.loni.usc.edu/wp-content/uploads/how\_to\_apply/ADNI\_Acknowledgement\_List.pdf
}
}

\maketitle

\begin{abstract}
Content-based image retrieval (CBIR) systems are an emerging technology that supports reading and interpreting medical images. Since 3D brain MR images are high dimensional, dimensionality reduction is necessary for CBIR using machine learning techniques. In addition, for a reliable CBIR system, each dimension in the resulting low-dimensional representation must be associated with a neurologically interpretable region. We propose a localized variational autoencoder (Loc-VAE) that provides neuroanatomically interpretable low-dimensional representation from 3D brain MR images for clinical CBIR. Loc-VAE is based on $\boldsymbol{\beta}$-VAE with the additional constraint that each dimension of the low-dimensional representation corresponds to a local region of the brain. The proposed Loc-VAE is capable of acquiring representation that preserves disease features and is highly localized, even under high-dimensional compression ratios (4096:1). The low-dimensional representation obtained by Loc-VAE improved the locality measure of each dimension by 4.61 points compared to na\"{i}ve $\boldsymbol{\beta}$-VAE, while maintaining comparable brain reconstruction capability and information about the diagnosis of Alzheimer's disease.
\end{abstract}


\section{Introduction}
Magnetic resonance (MR) images are stored in the picture archiving and communication system (PACS) \cite{choplin1992picture} along with the corresponding clinical information, which enables the centralized management of scanned images. These stored images are retrieved for diagnostic and research purposes. When querying and registering images in such databases, it is common to use keywords that describe brain structural and clinical features and so on. However, selecting the appropriate keywords requires sufficient experience in the specialized field. Therefore, it is desirable to develop a content-based image retrieval (CBIR) \cite{kumar2013content} system in medical practice to retrieve MR images by querying the images themselves rather than keywords.
\newcounter{num}

Since MR images are usually composed of millions of voxels or more, CBIR based on machine learning techniques must avoid the curse of dimensionality. Classic and widely used methods for dimensionality reduction for this purpose can be mainly categorized into two groups: (\setcounter{num}{1}\roman{num}) feature extraction, which transforms the part of the interest in the data into a compact vector \cite{huang2012retrieval,huang2014content,arakeri2013intelligent}, and (\setcounter{num}{2}\roman{num}) compressed expression acquisition, which converts the entire data into a vector of summaries with such as singular value decomposition \cite{lyra2012improved} or other means. However, feature extraction to obtain compact vectors generally is not easy and requires specialized feature engineering. However, acquiring compressed representation is challenging to balance low dimensionality and preservation of important features. With the recent advancement of deep learning-based techniques in computer vision, convolutional neural networks (CNNs) that can encompass (\setcounter{num}{1}\roman{num}) and (\setcounter{num}{2}\roman{num}) have been proposed, and have been applied to brain MR images \cite{dou2016automatic, esmaeilzadeh2018end, korolev2017residual}. In addition, several CNN-based algorithms for CBIR have been proposed \cite{owais2019effective,swati2019content,arai2018significant,onga2019efficient}. Especially, convolutional autoencoder (CAE)-based dimensionality reduction methods \cite{arai2018significant} have achieved a high compression ratio in brain MR images. Moreover, an extension 3D-CAE also utilizes metric learning to acquire more disease-specific low-dimensional representation \cite{onga2019efficient}.

\begin{figure*}[t]
    \centering
    \includegraphics[width=0.99\linewidth]{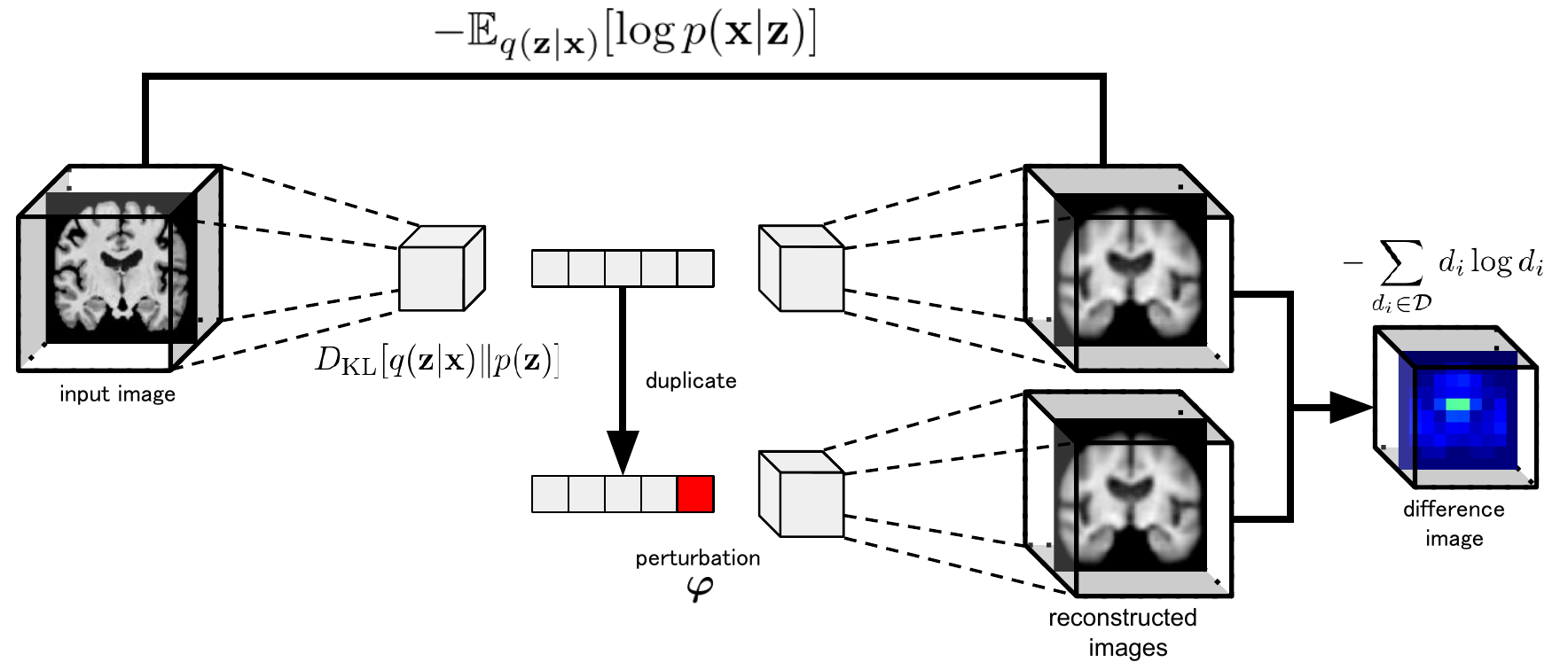}
    \caption{The schematics of the proposed method.}
    \label{fig:schematics}
\end{figure*}

To achieve a reliable CBIR system to support clinical decisions, users must be provided with human-interpretable reasons for the similarity of the images. However, CNN or CAE-based dimensionality reduction methods above do not consider the readability and interpretability of the obtained low-dimensional representation. For the CBIR system to enable image retrieval based on disease-related neuropathological features, each dimension in the resulting representation must be associated with a neurologically interpretable region containing known disease-related pathology. With these capabilities, such a CBIR system does not merely list the results but also provides the radiologist with the rationale for the system's recommendations, offering the possibility of using the results even more effectively.

This paper proposes a general-purpose, highly interpretable low-dimensional representation acquisition method, localized variational autoencoder (Loc-VAE), and applies it to brain MR images to implement a practical CBIR system. The proposed Loc-VAE adds a new constraint to the $\beta$-variational autoencoder ($\beta$-VAE) \cite{higgins2016beta}, resulting in a highly interpretable low-dimensional representation in which each dimension is independent and responsible for a specific portion of the input data, i.e., a local brain region.

\section{Related Works}
In this section, we mainly focus on the properties of the two CNN-based dimensionality reduction methods for CBIR and the interpretability of $\beta$-VAE.

Swati \cite{swati2019content} et al. proposed a framework for CBIR using VGG19 \cite{simonyan2014very} pre-trained on ImageNet \cite{russakovsky2015imagenet} and closed-form metric learning (CFML) \cite{alipanahi2008distance} of the similarity distance. The pre-trained VGG19 is fine-tuned on brain MR images with metric learning, used to determine the optimum metric, which increases intraclass similarity while decreasing interclass similarity. Similarity cases are determined by calculating the similarity between query and database images by applying CFML on features of the FC7 layer in VGG19. Swati et al.'s CNN-based model can acquire the features to find similarity cases without manually creating features. However, their model needs disease label information in their fine-tuning stage. Since CBIR is expected to include a variety of cases, it is not reasonable to build a model based on a classification task for all labels. The low-dimensional representation also retains features useful for classification, but the readability of each dimension is not mentioned.

Arai et al. proposed CAE-based dimensionality reduction in CBIR \cite{arai2018significant}. CAE is an extension of an autoencoder that uses a using CNN for compression and restoration. The basic idea behind dimensionality reduction with a CAE is that if the reconstruction error between the input and the output is small, the low-dimensional representation retains a large amount of input information. This methodology is practical in CBIR because the model can be trained without specific label information. Arai et al. have successfully compressed brain MR images of 5 million dimensions down to 150 dimensions while preserving clinically relevant neuroradiological features. Although a CAE provides a high compression performance by learning to reduce reconstruction errors, the image and its low-dimensional representation can be obtained only as a point-to-point relationship in the respective data space, and continuity around a data point is not guaranteed. In addition, the interpretability of the low-dimensional representation is not taken into considered.

Higgins et al. proposed $\beta$-VAE \cite{higgins2016beta}, a deep unsupervised generative approach for disentangled low-dimensional representation. Like a CAE, $\beta$-VAE is a CNN-based encoder-decoder model, with the most significant difference being that it assumes that the input data are generated from multivariate normal distributions. The encoder of $\beta$-VAE converts the input data into a low-dimensional probability distribution where each dimension follows normal distribution, and the decoder reconstructs the original data from the distribution. In other words, in $\beta$-VAE, a single data point is embedded as a low-dimensional probability distribution. Thus, unlike the CAE, $\beta$-VAE guarantees continuity around data points, so data that are close in the input space are expected to be placed close in lower-dimensional space. Moreover, since each dimension of the distribution is independent and regularized, the resulting low-dimensional representation is much more neuroanatomically interpretable than in the CAE case. These are important features in CBIR realization. However, few studies have obtained disentangled representation evaluated against brain MR images.

\section{Proposed Method}
In this paper, we propose the localized variational autoencoder (Loc-VAE), an encoder for acquiring interpretable low-dimensional representation from brain MR images for CBIR. Fig. \ref{fig:schematics} shows an overview of the proposed Loc-VAE. Loc-VAE is a learning model based on $\beta$-VAE \cite{higgins2016beta}, which provides independent embedding for each dimension while ensuring continuity for each localized region of the brain. The loss function of Loc-VAE consists of the following two terms:
\begin{equation}
    \mathcal{L}=\mathcal{L}_{\mathrm{\beta \text{-} VAE}}+\mathcal{L}_{\mathrm{Local}}.
\end{equation}
The first term, $\mathcal{L}_{\mathrm{\beta \text{-} VAE}}$, is the term used in general VAE models, and the second term, $\mathcal{L}_{\mathrm{Local}}$, is a newly introduced term to localize the range carried by each dimension of the low-dimensional representation and is related to the geometric variation of the output obtained by the decoder.

\subsection{$\beta$-VAE}
First, we briefly introduce $\beta$-VAE, the foundation of the proposed Loc-VAE. $\beta$-VAE consists of an encoder and a decoder, and is a generative model that estimates the generation probability $p(\mathbf{x})$ of the input data $\mathbf{x} \in \mathbb{R}^{d}$. Since it is inherently difficult to estimate $p(\mathbf{x})$ directly from the observation of $\mathbf{x}$, $\beta$-VAE accomplishes this by maximizing the conditional probability $p(\mathbf{x}|\mathbf{z})$, where $\mathbf{z} \in \mathbb{R}^{d'}$ is a latent variable (usually $(d \gg d’)$) obtained by its encoder. We assume that z is generated from a normal distribution $p(\mathbf{z})$, and let $q(\mathbf{z}|\mathbf{x})$ be the conditional probability distribution for estimating the distribution of $\mathbf{z}$. The target loss function of $\beta$-VAE to be minimized is defined by
\begin{equation}
    \mathcal{L}_{\mathrm{\beta \text{-} VAE}} = -\mathbb{E}_{q(\mathbf{z}|\mathbf{x})}[\log p(\mathbf{x}|\mathbf{z})] + \beta D_{\mathrm{KL}}[q(\mathbf{z}|\mathbf{x}) \| p(\mathbf{z})].
    \label{eq:vae}
\end{equation}
The first term is the reconstruction error between the input and the reconstructed output of the data (i.e., brain MR images), the second term is Kullback–Leibler (KL) divergence between the estimated generating distribution for $\mathbf{z}$ obtained by the encoder and the normal distribution $p(\mathbf{z})$, and $\beta$ is a hyperparameter that balances the two terms.

The latent variable $\mathbf{z}$ is sampled from the generated distribution $q(\mathbf{z}|\mathbf{x})$, using the mean $\boldsymbol{\mu}$ and the variance $\boldsymbol{\sigma}$ obtained by the encoder. However, since sampling is not a continuous computation, backpropagation cannot be used to acquire the parameters associated with it. Therefore, for VAEs involving $\beta$-VAE, the reparameterization trick is used to approximate it as
\begin{equation}
    \mathbf{z} = \boldsymbol{\mu} + \boldsymbol{\epsilon} \odot \boldsymbol{\sigma}, \quad \boldsymbol{\epsilon} \sim N(0, \mathbf{I}).
    \label{eq:repara}
\end{equation}
The reparameterization trick calculates the latent variables using perturbations $\boldsymbol{\epsilon}$ sampled from $N(0, \mathbf{I})$, where $\odot$ is an element-wise product.

\eqref{eq:vae} and \eqref{eq:repara} allow $\beta$-VAE to acquire the highly independent low-dimensional latent representation $\mathbf{z}$ (more precisely, their distribution) for each dimension while retaining much of the input information. Since each dimension of the latent distribution $q(\mathbf{z}|\mathbf{x})$ is highly independent and normalized so that the mean is zero and the variance is close to one, it is easier to interpret them than the low-dimensional representation obtained by the CAE. This feature is also desirable for later machine learning processing. Furthermore, as mentioned above, $\beta$-VAE embeds data into a probability distribution, ensuring continuity in the vicinity of data points, making it suitable for CBIR realization.

\subsection{The Local Loss}
The key element of Loc-VAE, the local loss, is a constraint that localizes the influence of each dimension of the latent variable $\mathbf{z}$. Specifically, we first obtain two reconstructions, one with and one without the one-hot-vector perturbations $\boldsymbol{\varphi}$, which is non-zero in only one dimension randomly selected for the embedding $\mathbf{z}$. Then, we obtain the difference between the two reconstructed images $\mathcal{D}$ as follows\footnote{The equations in the text are strictly written, and an intuitive expression that follows the notation used in a typical VAE would be as follows: $\mathcal{D} = \mathbb{E}_{q(\mathbf{z}|\mathbf{x})}\left[|p(\mathbf{x}|\mathbf{z})-p(\mathbf{x'}|\mathbf{z}+\boldsymbol{\varphi})|\right]$.}:
\begin{equation}
    \mathcal{D} = \mathbb{E}_{q(\mathbf{z} | \mathbf{x})}\left[\mathbb{E}_{p(\mathbf{x} | \mathbf{z}), p(\mathbf{x'} | \mathbf{z}+\boldsymbol{\varphi})}[|\mathbf{x}-\mathbf{x'}|]\right].
\end{equation}
This means the extent to which a given dimension affects the reconstructed image. Each pixel in the reconstructed image is considered a probability distribution with a dimension of the number of pixels, normalized by the maximum and minimum values of the corresponding pixel in the entire dataset. Local loss is defined as the entropy of the difference image $\mathcal{D}$ with each pixel $d_i$, the amplitude of the perturbation $\boldsymbol{\varphi}$, and $\gamma$ is a hyperparameter:
\begin{equation}
    \mathcal{L}_{\mathrm{Local}}=-\gamma\sum_{d_i \in \mathcal{D}} d_i \log d_i.
    \label{eq:local}
\end{equation}
In other words, $\mathcal{L}_{\mathrm{Local}}$ is large when each dimension of $\mathbf{z}$ is a representation that affects a wide area of the image. This allows learning to occur so that each dimension of the low-dimensional representation has a narrowed range of influence on the corresponding brain MR images and is expected to obtain localized low-dimensional representation.

\begin{figure}[t]
    \centering
    \begin{minipage}{0.3\linewidth}
        \includegraphics[width=\linewidth]{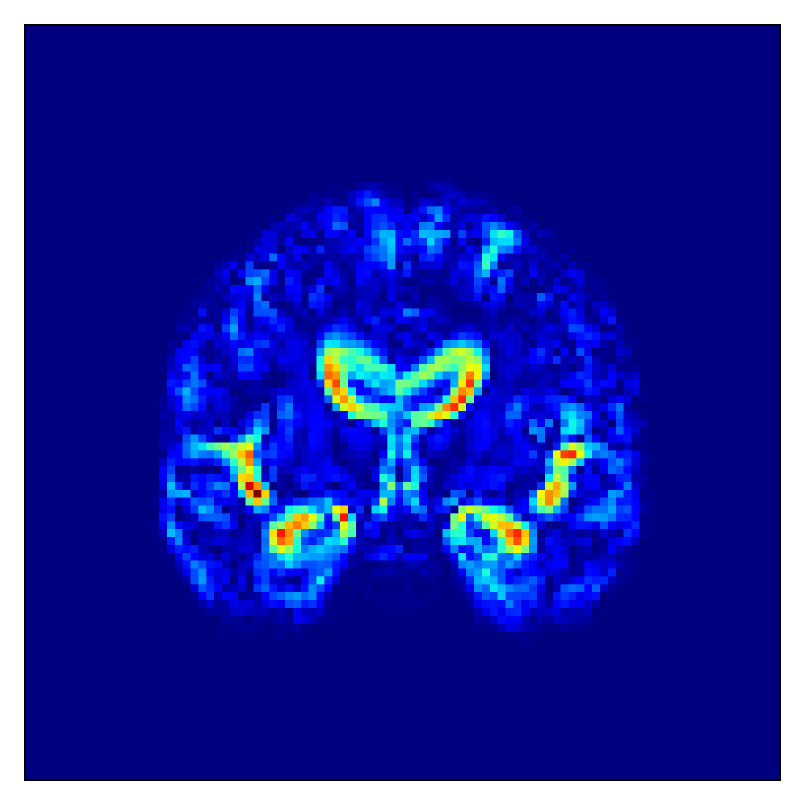}
        \subcaption{Coronal}
    \end{minipage}
    \begin{minipage}{0.3\linewidth}
        \includegraphics[width=\linewidth]{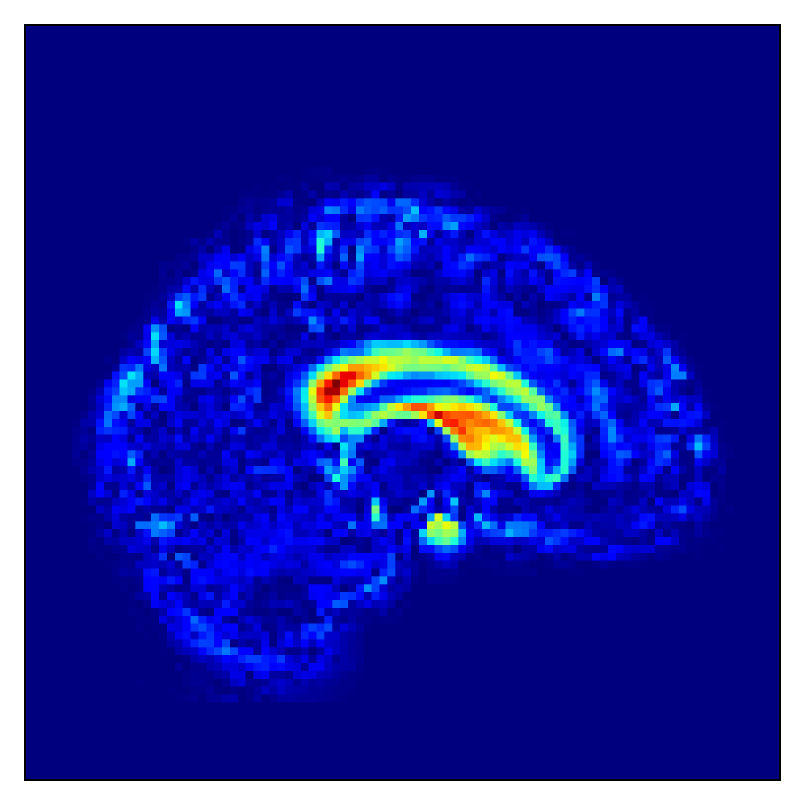}
        \subcaption{Sagittal}
    \end{minipage}
    \begin{minipage}{0.3\linewidth}
        \includegraphics[width=\linewidth]{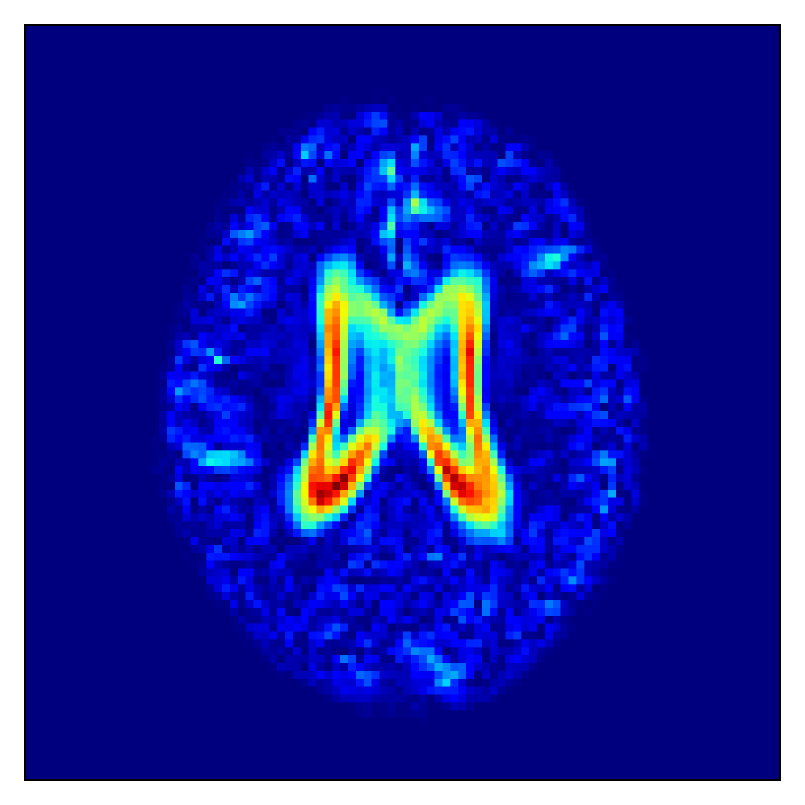}
        \subcaption{Transverse}
    \end{minipage}
   \caption{The difference between the mean images of the AD and CN conditions.}
   \label{fig:mean}
\end{figure}

\begin{figure*}[t]
    \centering
    \includegraphics[width=\linewidth]{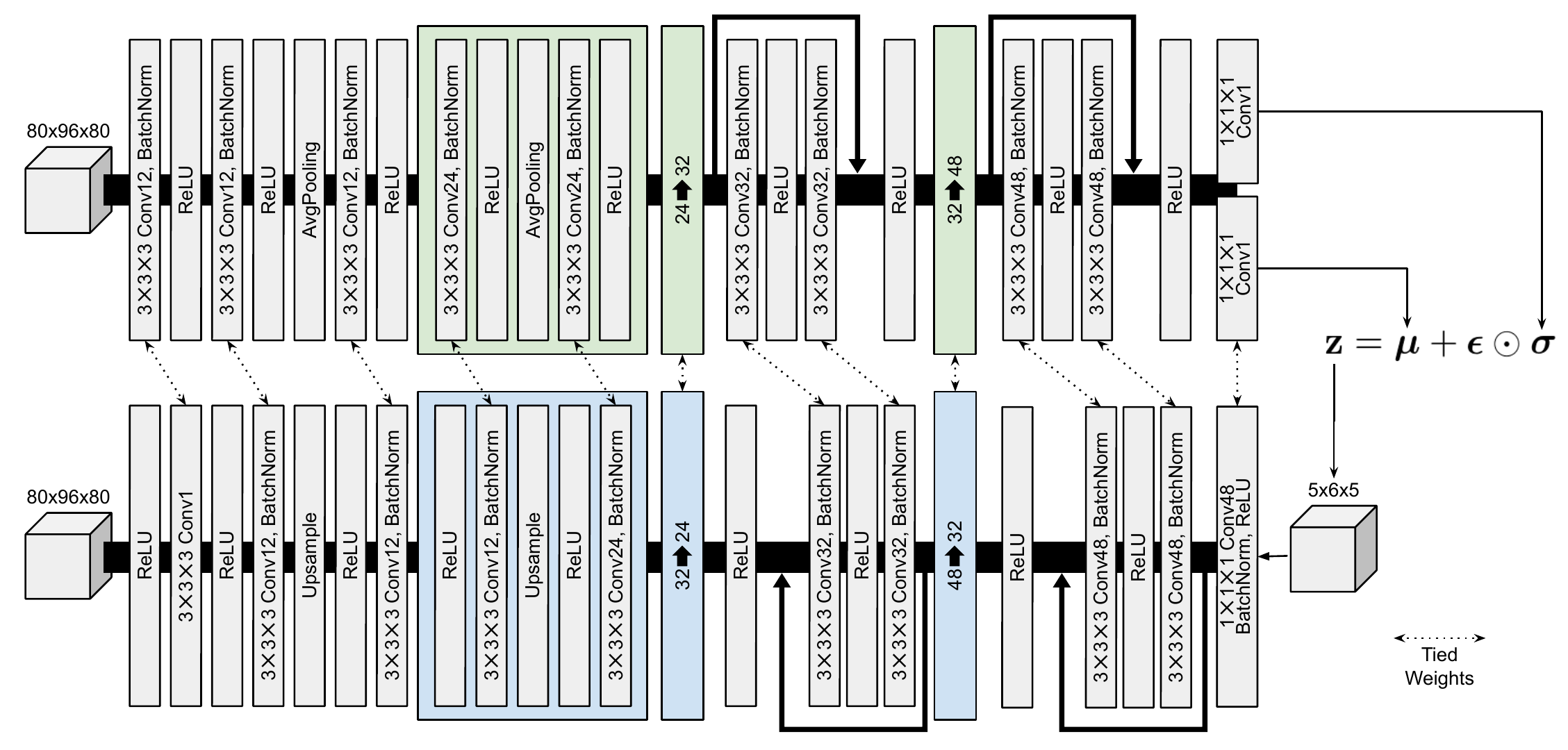}
    \caption{Network architecture of Loc-VAE in this experiment. Low-dimensional representation $\mathbf{z}$ is calculated using the obtained $\boldsymbol{\mu}$ and $\boldsymbol{\sigma}$ through the reparameterization trick.}
    \label{fig:network}
\end{figure*}

\section{Experiments}
In CBIR, as mentioned, acquired low-dimensional representation should retain the original brain information and be interpretable; that is, each of those dimensions should have its structural information correspond to a local region. Therefore, in this experiment, we evaluated the low-dimensional representation obtained with the proposed method in terms of the reconstruction performance (i.e., information preservation), the extent to which each dimension is affected (i.e., readability), and the diagnostic capability based on the representation (i.e., applicability of the representation), and compared them with $\beta$-VAE as a performance baseline.

\subsection{Dataset and Preprocessing}
In this experiment, we used T1-weighted brain MR images with disease information from the Alzheimer's Disease Neuroimaging Initiative 2 (ADNI2)\footnote{https://adni.loni.usc.edu} dataset. Four categories of data were used in the experiment: cognitively normal (CN), mild cognitive impairment (MCI), late-MCI (LMCI), and Alzheimer's disease (AD). The last is the worst condition. All images were skull-stripped and linearly aligned to the JHU-MNI space \cite{oishi2009atlas} using MRICloud \cite{mori2016mricloud}\footnote{https://mricloud.org/} as the preprocessing method, resulting in aligned brain images. MRICloud provides excellent skull-stripping performance, yielding adequate brain regions in most cases, but rarely fails. An experienced neurologist performed quality control of the MR images and excluded MR images that were not properly preprocessed.

After the background region was cropped, skull-stripped brain images were downsampled into $80 \times 96 \times 80$ voxels with bicubic interpolation. Pixels in each image with pixel values less than $0$ or greater than four times the standard deviation were excluded as outliers, and then linear min-max normalization was conducted. The excluded pixels were replaced with the minimum and maximum values after normalization. The final number of MR images used in the experiments was $3,716$ from $1,170$ subjects, including $360$ CN, $240$ MCI, $280$ LMCI, and $290$ AD cases. As described in detail below, cases from the same patient were not split into training and evaluation for the model. This was to prevent bias caused by mixing similar data into training and validation fold.

Fig. \ref{fig:mean} shows one slice of the difference image between the CN and AD mean images in the ADNI database. Alzheimer's features are located primarily near the ventricles. To assess similarity cases based on disease features, it is necessary that one dimension of the low-dimensional representation captures this region.

\subsection{Detail of the model}
Fig. \ref{fig:network} shows the architectural diagram of the proposed Loc-VAE used in this experiment. Since Loc-VAE is a model that adds the local loss term shown in \eqref{eq:local} to $\beta$-VAE, the structure of these models is the same and can be chosen arbitrarily. In the proposed model, the encoder consists of $13$ convolutional layers, average pooling, and skip connections \cite{he2016deep}. In the experiments, as in previous studies \cite{arai2018significant}, latent variable $\mathbf{z}$ was set to $150$ dimensions. Specifically, the final layer of the encoder was split to obtain the mean $\boldsymbol{\mu}$ and the variance $\boldsymbol{\sigma}$, which are two $5 \times 6 \times 5$ (i.e., 150) dimensional outputs. The resulting dimensionality compression ratio was ($80 \times 96 \times 80$):$150$, or $4,096$:$1$.

The design of the decoder is in contrast to the encoder by using convolutional layer and upsampling. To prevent the obtained latent representation $\mathbf{z}$ from being overestimated to be localized due to the decoder configuration, Loc-VAE introduced tied weights or weight tying \cite{vincent2010stacked}, where the encoder and decoder share weights.

\subsection{Evaluation}
We assessed the effectiveness of the proposed Loc-VAE using the following three perspectives and compared it to $\beta$-VAE as the baseline:
\begin{enumerate}[(a)]
 \item Accuracy of MR image reconstruction
 \item Locality of each low-dimensional representation
 \item Classification performance.
\end{enumerate}

In (a), reconstruction errors were evaluated to assess the extent to which the resulting low-dimensional representation preserved the structural information of the brain MR images. We used root mean squared error (RMSE) and structured similarity (SSIM) criteria.

In (b), to assess the readability of the representation, we investigated their locality in each dimension. Specifically, two reconstructed images were created by adding perturbations of $+3\sigma$ and $-3\sigma$ to the low-dimensional representation z for one specific dimension, respectively. The difference image between these two images was considered a probability distribution, and the entropy was calculated.

In (c), to assess the applicability of the representation for CBIR, we investigated the classification performance of that representation. We trained a logistic regression model to classify AD and CN from their low-dimensional representation and evaluated their performance in terms of the area under the receiver operating characteristics (AUC) curve.

Since the VAE model contains the reparametrization trick that simulates a non-deterministic sampling process, we took the average of 10 trials for each evaluation. In the experiments, all models were trained and evaluated using group $5$-fold cross-validation to avoid mixing the same patient in the training and validation data. In other words, cases of the same patient are not split into training and evaluation data for the model.

\section{RESULTS}
\begin{figure}[t]
    \centering
    \begin{minipage}{0.43\linewidth}
        \includegraphics[width=\linewidth]{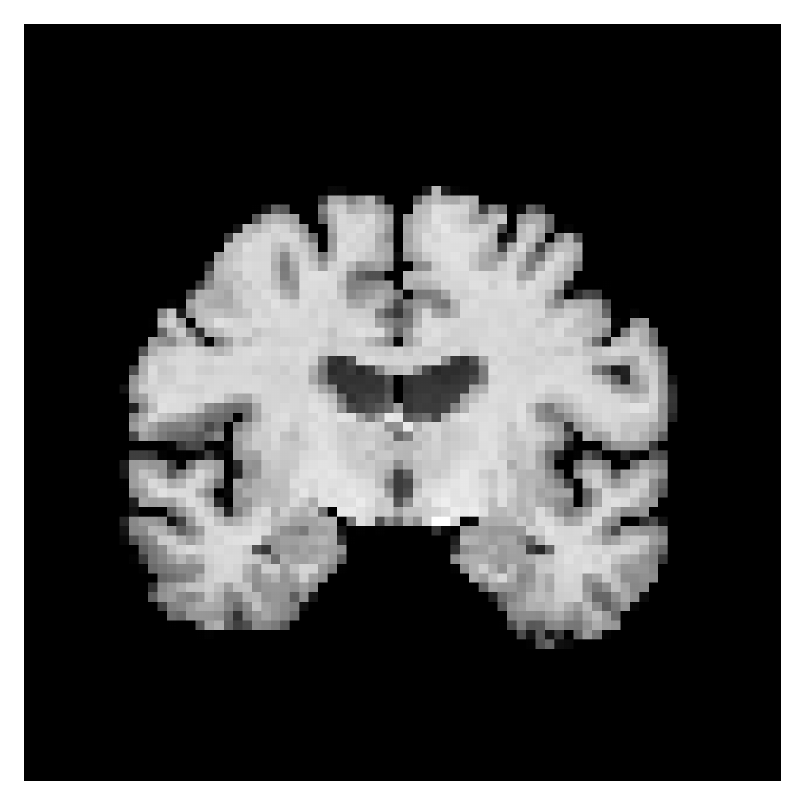}
        \subcaption{Ground truth}
    \end{minipage}
    \begin{minipage}{0.43\linewidth}
        \includegraphics[width=\linewidth]{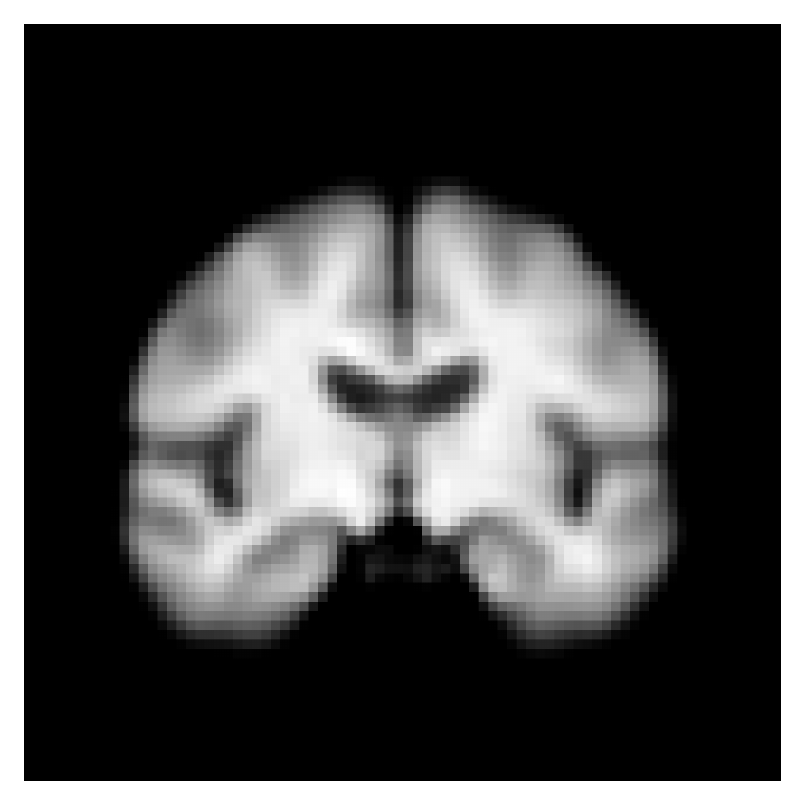}
        \subcaption{$\beta$-VAE}
    \end{minipage} \\
    \begin{minipage}{0.43\linewidth}
        \includegraphics[width=\linewidth]{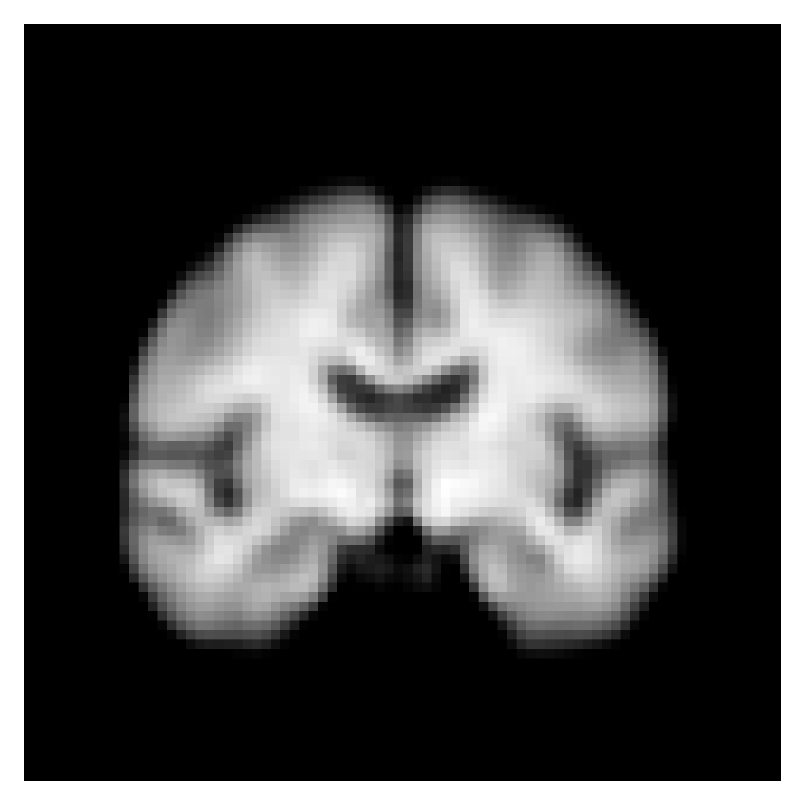}
        \subcaption{$\beta$-VAE (TW)}
    \end{minipage}
    \begin{minipage}{0.43\linewidth}
        \includegraphics[width=\linewidth]{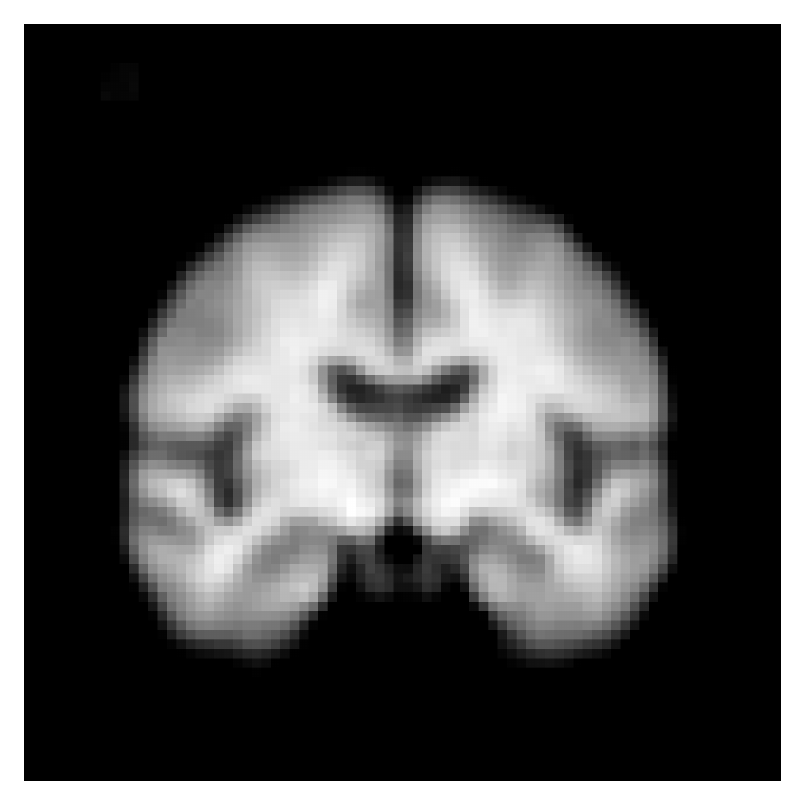}
        \subcaption{Loc-VAE}
   \end{minipage}
   \caption{Reconstructed brain MR images (the coronal plane).}
   \label{fig:reconst}
\end{figure}

\begin{table}[t]
    \centering
    \caption{Comparison of Model Performance}
    \label{table:result}
    \begin{tabular}{@{}lrrrr@{}}
    \toprule
    \multicolumn{1}{c}{} & \multicolumn{1}{c}{\textbf{RMSE}} & \multicolumn{1}{c}{\textbf{SSIM}} & \multicolumn{1}{c}{\textbf{ENTROPY}} & \multicolumn{1}{c}{\textbf{AUC}} \\ \midrule
    $\beta$-VAE                & 0.0947                   & 0.827                    & 6.55                        & 0.739                   \\
    $\beta$-VAE (TW)           & 0.0956                   & 0.822                    & 6.65                        & 0.740                   \\
    Loc-VAE        & 0.0974                   & 0.815                    & 2.04                        & 0.718                   \\ \bottomrule
    \end{tabular}
\end{table}

\begin{figure}[t]
    \centering
    \begin{tabular}{ccc}
    \includegraphics[width=0.3 \linewidth]{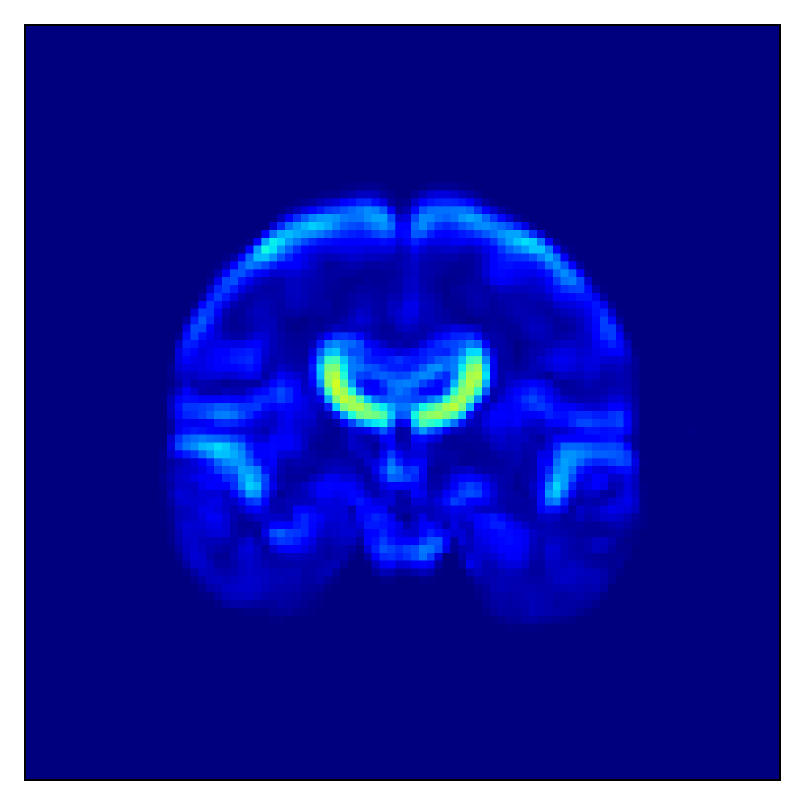}&\includegraphics[width=0.3\linewidth]{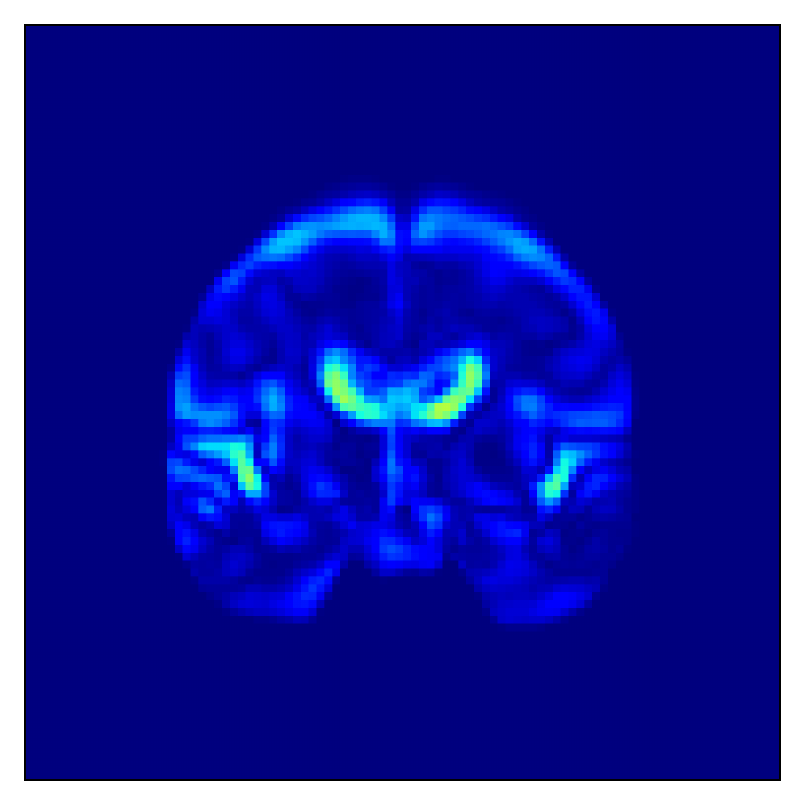}&\includegraphics[width=0.3 \linewidth]{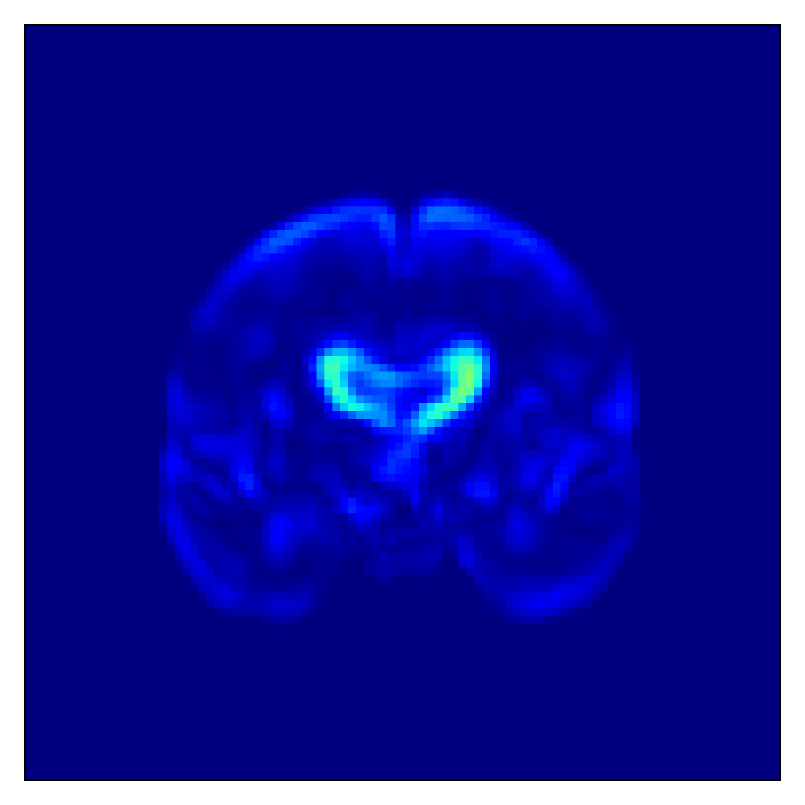}\\
    \includegraphics[width=0.3 \linewidth]{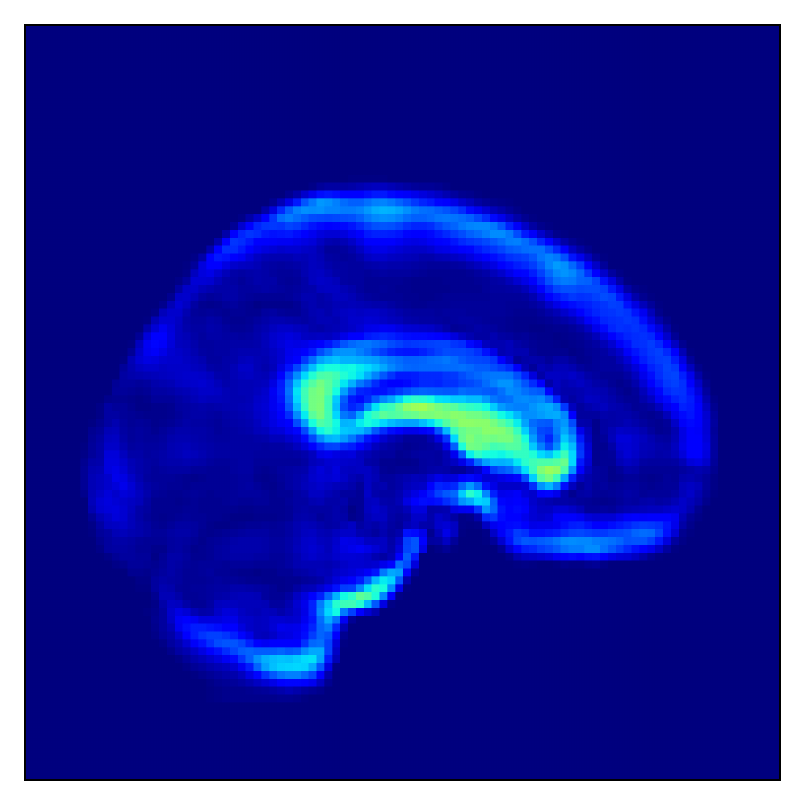}&\includegraphics[width=0.3 \linewidth]{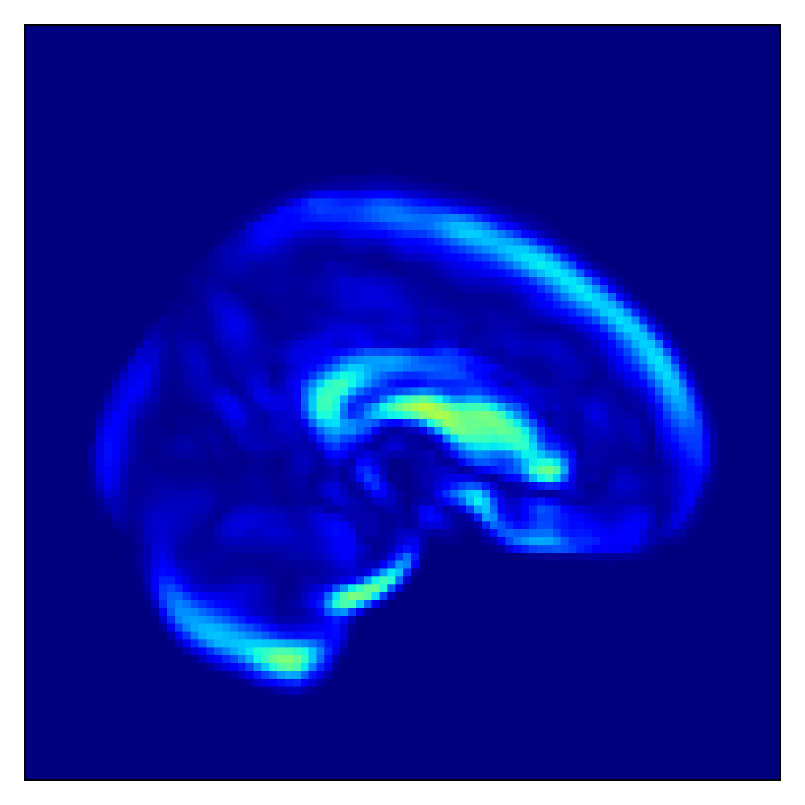}&\includegraphics[width=0.3 \linewidth]{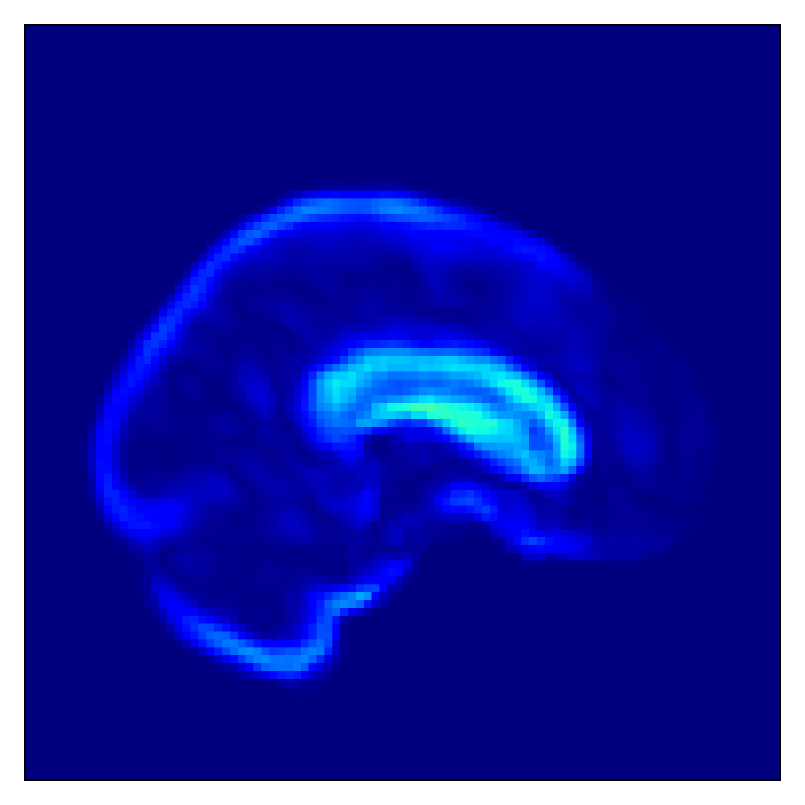}\\
    \includegraphics[width=0.3 \linewidth]{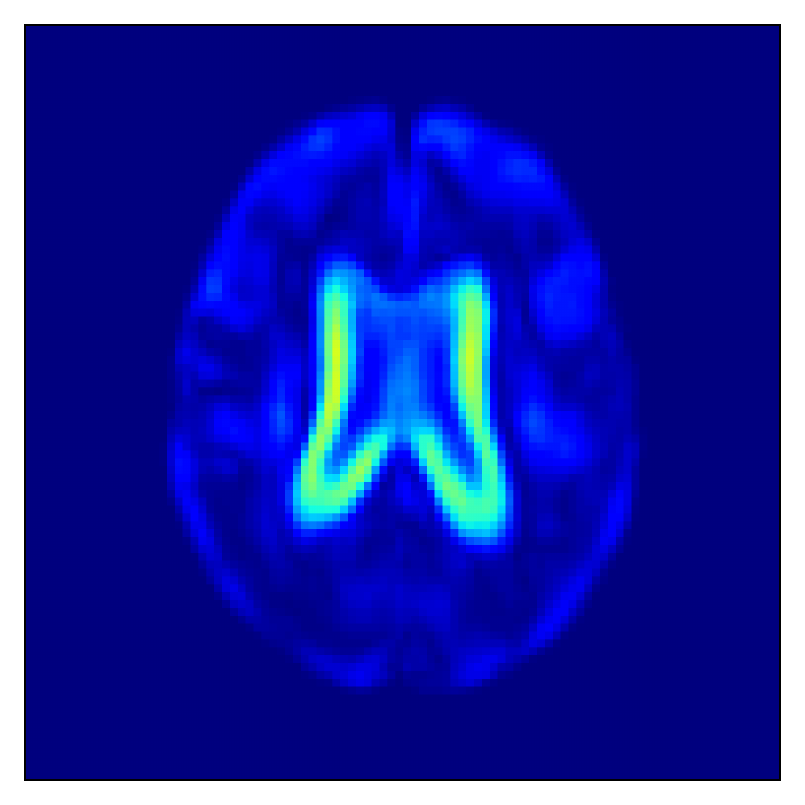}&\includegraphics[width=0.3 \linewidth]{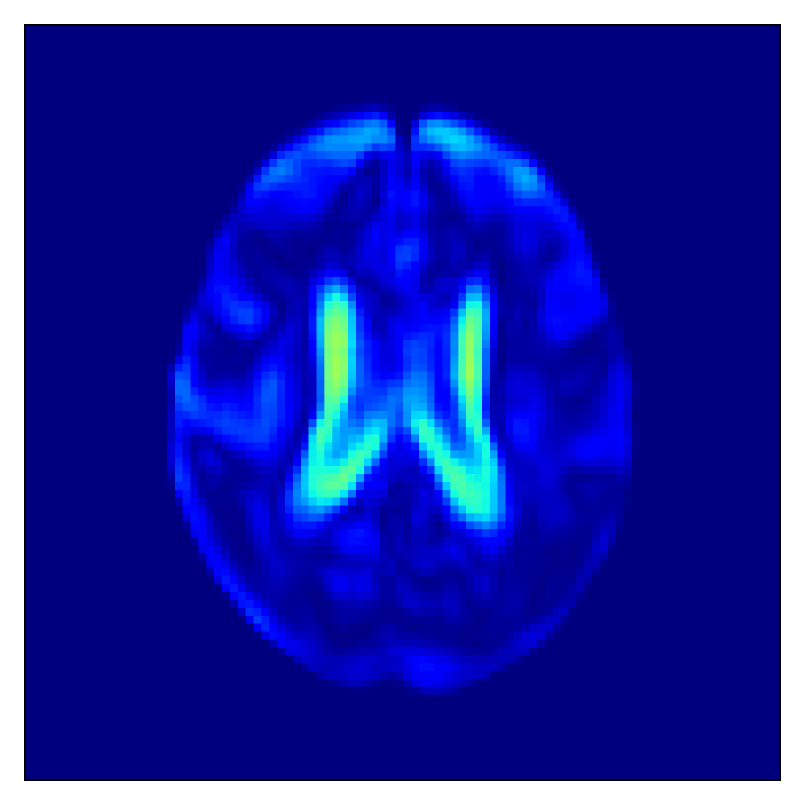}&\includegraphics[width=0.3 \linewidth]{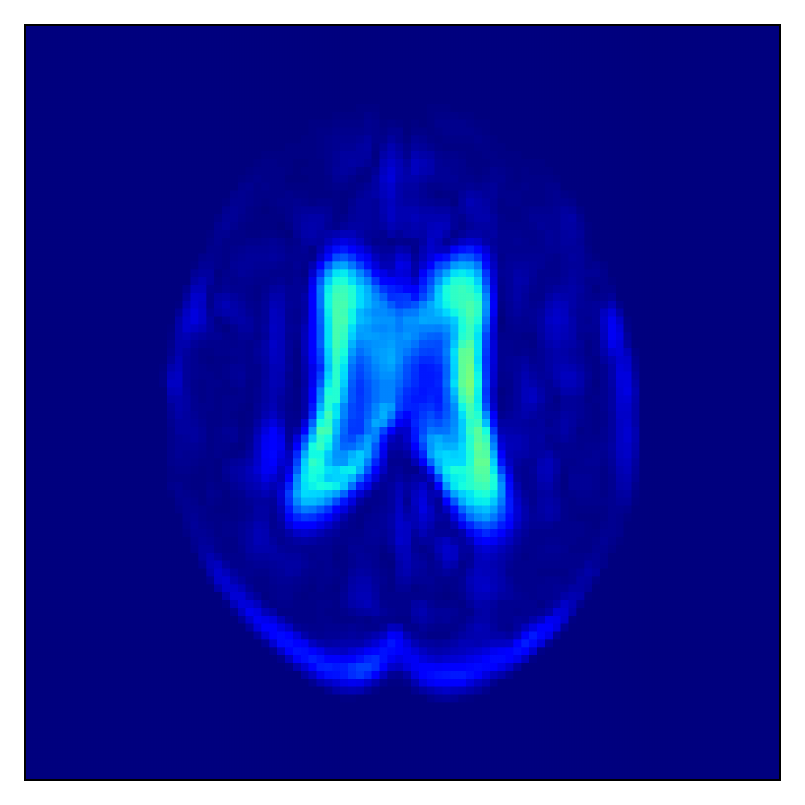}\\  
    (a) & (b) & (c)
    \end{tabular}
    \caption{Brain regions affected by the one dimension with the largest difference between the AD and CN conditions among the low-dimensional representation obtained. (a) na\"{i}ve $\beta$-VAE, (b) $\beta$-VAE (TW), and (c) Loc-VAE.}
    \label{fig:localized}
\end{figure}

Table \ref{table:result} summarizes the performance comparison. Since the proposed Loc-VAE adopts tied weights, $\beta$-VAE with the tied weights ($\beta$-VAE (TW)) was included in the evaluation in addition to the na\"{i}ve $\beta$-VAE.

Fig. \ref{fig:reconst} shows an example of reconstruction (the coronal plane) by $\beta$-VAE, $\beta$-VAE with TW, and Loc-VAE. Note that it is the expected value of $10$ reparameterization tricks.

Fig. \ref{fig:localized} shows the difference in the reconstructed images (the coronal, sagittal, transverse planes) with and without perturbation to one dimension of the low-dimensional representations that most affect the disease features between AD and CN. This one dimension is chosen to have the largest expected value of the difference in the mean vector ($150$ dim) between AD and CN.

Fig. \ref{fig:boxplot} shows the box-and-whisker diagram of the entropy of the reconstructed image when each dimension of the low-dimensional representation is perturbed. In other words, it represents the size, or locality, of the area covered by information about each dimension. This result is based on a random selection of one of the five folds, with the other folds showing similar results.

\section{DISCUSSION}
\subsection{Accuracy of MR Image Reconstruction}
The proposed Loc-VAE reconstructs enough brain images to determine that the low-dimensional representation retains the structural information of the input brain. The reconstruction accuracy of Loc-VAE is equivalent to that of VAEs, and to the reconstruction capacity of a brain MR image of the same size compressed using a CAE to the same 150 dimensions \cite{arai2018significant}. Few negative effects due to the introduction of local loss were observed ($+0.0018$ in RMSE and $-0.007$ in SSIM) with respect to the brain image reconstruction.

\subsection{Locality for Each Low-dimensional Representation}
In Fig. \ref{fig:localized}, the influence of dimensions that may contribute to the diagnosis of AD is examined, and it can be seen that the na\"{i}ve $\beta$-VAE captures not only the important areas around the ventricles shown in Fig. \ref{fig:mean} but also the edges of the brain and other areas. Loc-VAE, however, is more limited and captures this region better. This result shows that Loc-VAE acquires a specific dimension of the disease features on low-dimensional representation. Disease feature--specific dimensions serve as materials for the neurologist to assess similar cases displayed by CBIR.

\begin{figure}[t]
    \centering
    \includegraphics[width=0.9\linewidth]{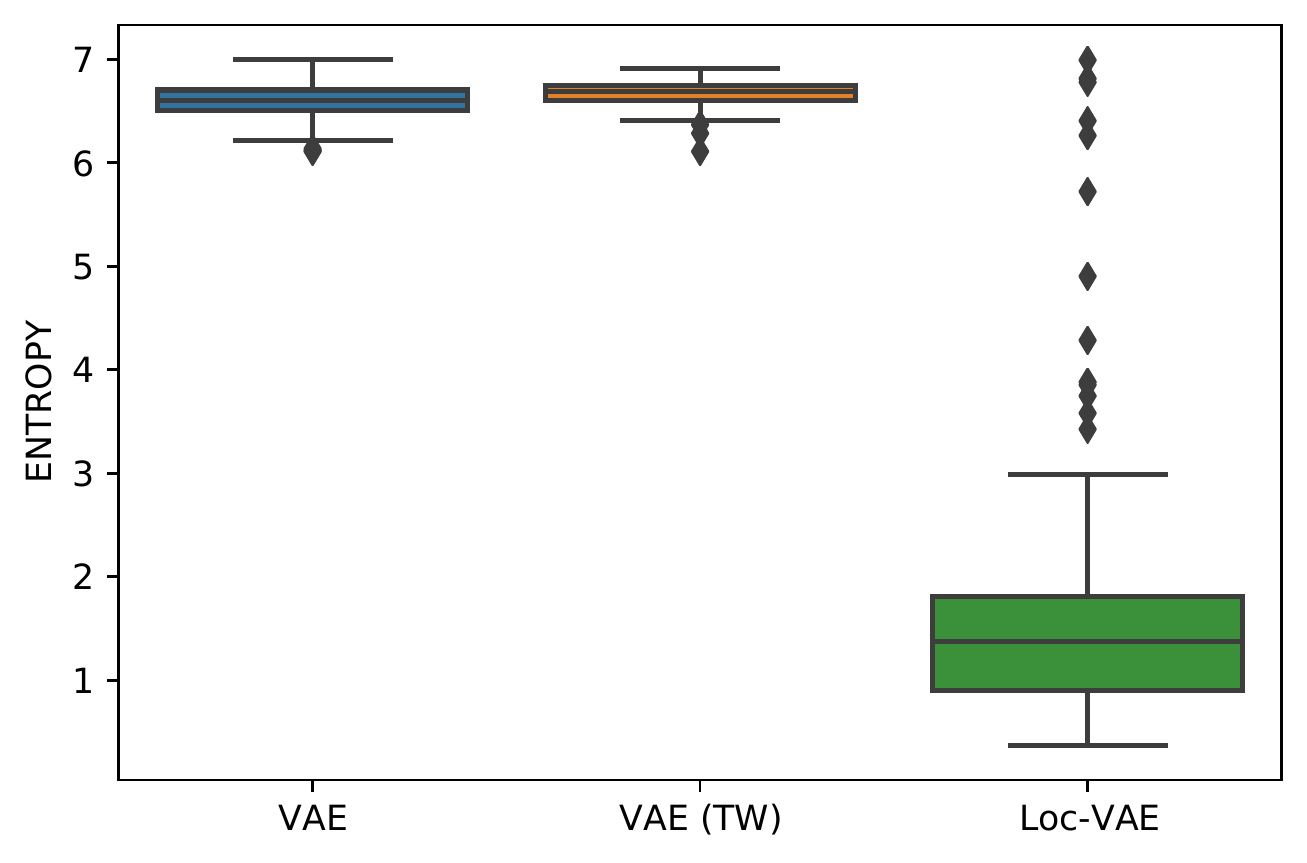}
    \caption{Box-and-whisker plot when perturbations are applied to each dimension of the low-dimensional representation.}
    \label{fig:boxplot}
\end{figure}

In addition, it is quantitatively confirmed in Fig. \ref{fig:boxplot} that each of the low-dimensional representation obtained with Loc-VAE covers a local range of brain structures, thanks to the introduction of local loss. This improvement in the readability of the low-dimensional representation is important for CBIR realization and is a major achievement of this study. 

\subsection{Classification Performance}
The diagnostic performance based on the low-dimensional representation obtained with Loc-VAE was slightly lower than that of $\beta$-VAE (--0.022 in the AUC) but in a range that could be considered equivalent. This result shows that Loc-VAE performs dimensionality reduction while preserving disease information. This study aims not to diagnose disease directly but to obtain an interpretable low-dimensional representation that retains as much of the original brain information as possible. The proposed Loc-VAE was able to acquire a representation that achieves these desirable features for CBIR realization.

\section{CONCLUSION}
In this paper, we proposed Loc-VAE, which provides an interpretable low-dimensional representation from 3D brain MR images for clinical CBIR. Loc-VAE is based on $\beta$-VAE with the additional constraint that each dimension of the low-dimensional representation of the brain that can be obtained has a local influence. The representation obtained by Loc-VAE retained well the information about the original brain structure and disease, and each of its dimensions covered a local area; that is, it was a highly readable and desirable representation. These features are important for CBIR realization, and the experiments demonstrated promising results. In the future, we will verify Loc-VAE on a larger scale.

\section{Acknowledgments}
\scriptsize{This research was supported in part by the Ministry of Education, Science, Sports and Culture of Japan (JSPS KAKENHI), Grant-in-Aid for Scientific Research (C), 21K12656, 2021-2023.

The MRI data collection and sharing for this project was funded by the Alzheimer's Disease Neuroimaging Initiative (ADNI) (National Institutes of Health Grant U01 AG024904) and DOD ADNI (Department of Defense award number W81XWH-12-2-0012). ADNI is funded by the National Institute on Aging, the National Institute of Biomedical Imaging and Bioengineering, and through generous contributions from the following: AbbVie, Alzheimer's Association; Alzheimer's Drug Discovery Foundation; Araclon Biotech; BioClinica, Inc.; Biogen; Bristol-Myers Squibb Company; CereSpir, Inc.; Cogstate; Eisai Inc.; Elan Pharmaceuticals, Inc.; Eli Lilly and Company; EuroImmun; F. Hoffmann-La Roche Ltd and its affiliated company Genentech, Inc.; Fujirebio; GE Healthcare; IXICO Ltd.; Janssen Alzheimer Immunotherapy Research \& Development, LLC.; Johnson \& Johnson Pharmaceutical Research \& Development LLC.; Lumosity; Lundbeck; Merck \& Co., Inc.; Meso Scale Diagnostics, LLC.; NeuroRx Research; Neurotrack Technologies; Novartis Pharmaceuticals Corporation; Pfizer Inc.; Piramal Imaging; Servier; Takeda Pharmaceutical Company; and Transition Therapeutics. The Canadian Institutes of Health Research is providing funds to support ADNI clinical sites in Canada. Private sector contributions are facilitated by the Foundation for the National Institutes of Health (www.fnih.org). The grantee organization is the Northern California Institute for Research and Education, and the study is coordinated by the Alzheimer's Therapeutic Research Institute at the University of Southern California. ADNI data are disseminated by the Laboratory for Neuro Imaging at the University of Southern California.
}


\bibliography{reference}
\bibliographystyle{IEEEtran}

\end{document}